\documentclass{llncs}
\usepackage{epsfig}

\begin{document}
\title{On Bus Graph Realizability\thanks{
Our interest in bus graph problems is a consequence
of the participation of one of us in Bertinoro Workshop on Graph Drawing, 5--10 March, 2006
where the problem was presented as an open problem.
Contact author: Ethan Kim,
School of Computer Science, McGill University, 3480 University St., Montr\'{e}al, Canada, \tt{ethan@cs.mcgill.ca}}}
\author{
Anil Ada \and
Melanie Coggan \and
Paul Di Marco \and
Alain Doyon \and
Liam Flookes \and
Samuli Heilala \and
Ethan Kim \and
Jonathan Li On Wing \and
Louis-Francois Preville-Ratelle\and
Sue Whitesides \and
Nuo Yu
}
\institute{School of Computer Science\\ 
  McGill University\\Montr\'{e}al, Canada
}

\maketitle

\begin{center}
\textbf{Technical Report SOCS-TR-2006.1, September 2006}
\end{center}
  
 
\begin{abstract}
  In this paper, we consider the following graph embedding problem:
  Given a bipartite graph $G=(V_1, V_2; E)$, where the maximum degree of vertices
  in $V_2$ is 4, can $G$ be embedded on a two dimensional grid such that 
  each vertex in $V_1$ is drawn as a line segment along a grid line, each
  vertex in $V_2$ is drawn as a point at a grid point, and each edge 
  $e=(u, v)$ for some $u\in V_1$ and $v\in V_2$ is drawn as a line segment 
  connecting $u$ and $v$, perpendicular to the line segment for $u$? 
  We show that this problem is NP-complete, and sketch how our proof techniques
  can be used to show the hardness of several other related problems.
\end{abstract}
 
\section{Introduction}
Orthogonal graph drawing is a well studied area in the graph drawing community, and 
one may find many applications in VLSI design. In this paper, we study orthogonal drawings
of \emph{bus graphs}, which represent interconnectivity of functional entities in a chip.
In VLSI layouts, a \emph{bus} is a line segment drawn on a plane. To establish a connection 
among a collection of buses, a \emph{connector} is drawn as a point on the plane, and then joined to the
buses by line segments. Thus, the interconnections of the buses can be represented by 
a bipartite graph, where one partition of vertices corresponds
to the set of buses, and the other corresponds to the set of connectors. 
We call such bipartite graphs \emph{bus graphs}, and we are interested in the problem of
drawing bus graphs on the plane.

For manufacturing purposes, it is desired that all the buses and the edges that join them are 
laid out either horizontally or vertically. Furthermore, it is also necessary to lay out the 
components of a chip, i.e. buses and connectors, sufficiently far from each other. Thus, we wish 
to draw the bus graph on a grid, where each bus is laid out along a grid line, and each 
connector is drawn at a grid point. Each edge between a bus and a connector is also drawn along 
a grid line. In particular, an edge joining a bus is drawn as a line segment perpendicular 
to the bus. We do not allow any edge to intersect with any connectors or buses, except at endpoints
of the edge. However, a horizontal edge may cross a vertical edge. Thus, each connector can 
connect at most 4 buses. We define the combinatorial bus graphs as the class of  bipartite 
graphs $G=(\mathcal{B},\mathcal{C};\mathcal{E})$, where the degree of connector vertices in 
$\mathcal{C}$ is at most 4. 

We say a (combinatorial) bus graph $G$ is \emph{realizable} if $G$ can be drawn while meeting the conditions above. 
See Fig.~\ref{fig:ab-perp} for an example of a bus graph and its realization.
We now formally define the Bus Graph Realizability (BGR) problem as follows:

\begin{figure}[tb]
\begin{center}
\epsfig{file=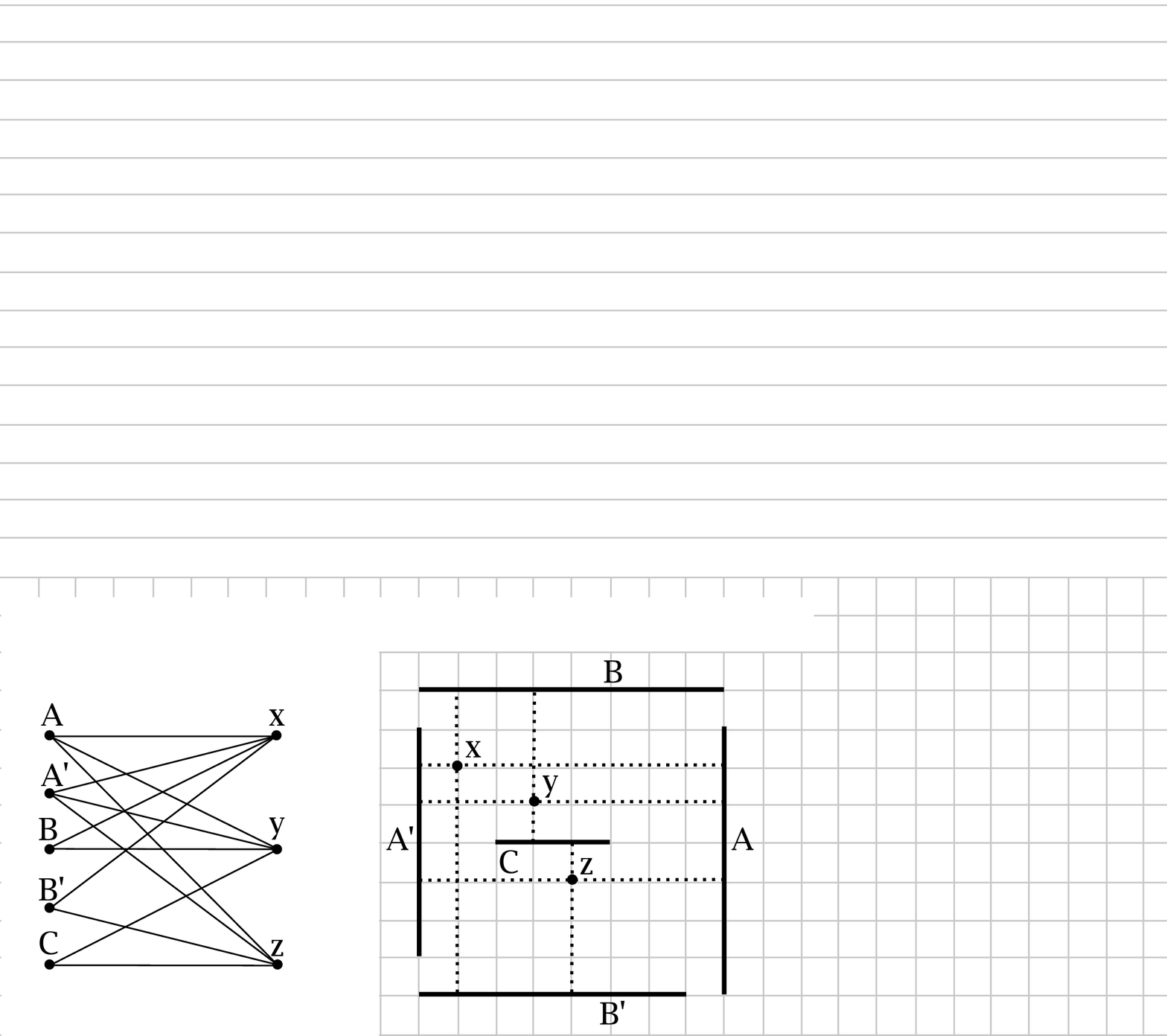, height=5cm, clip=}
\end{center}
\caption{A combinatorial bus graph and its realization. This graph is a gadget called an $(A, B)$-perp, 
  defined in Sect.~3.2, Definition~\ref{def:ab-perp}.}
\label{fig:ab-perp}
\end{figure}

\begin{description}
\item \textbf{Instance}: A bipartite graph $G=(\mathcal{B},\mathcal{C};\mathcal{E})$ 
  such that $\forall c\in \mathcal{C}$, $\deg(c)\leq 4$.
\item \textbf{Question}: Can $G$ be drawn onto a grid so that the following properties hold?
  \begin{enumerate}
    \item Each vertex $B \in \mathcal{B}$ is drawn as a closed line segment along a grid line.
    \item Each vertex $c \in \mathcal{C}$ is drawn as a point at a grid point.
    \item Each edge   $(B,c) \in \mathcal{E}$ is drawn as a closed line segment between $B$ and $c$, that
      is perpendicular to $B$, and contains no other connectors or buses apart from $B$ and $c$; an edge
      can, however, cross other edges as shown in Fig.~\ref{fig:ab-perp}. An edge may
      connect to a bus at an endpoint of the bus.
    \item No buses or connectors may intersect.
  \end{enumerate}
\end{description}
Now we are ready to state our main result.

\begin{theorem}
\label{thm1}
  Bus Graph Realizability is NP-complete.
\end{theorem}

The rest of this paper is organized as follows. In Sect. 2, we give definitions and
preliminaries. A proof of our main theorem is given in Sect. 3. In Sect. 4, we
show that the techniques we use to prove Theorem~\ref{thm1} can be applied to related problems.
In Sect. 5, we look at a variation of the problem where the lengths of the buses are given as input, 
and prove that this version of the problem is NP-hard.
Finally, we conclude with some open problems in Sect. 6.

\subsubsection{Related Work.} 
The orthogonal graph drawing style has found many applications in VLSI design since its
introduction in~\cite{thompson1979}, \cite{thompson1980}, and~\cite{bhatt1983}. 
Many optimization criteria have been suggested, such as minimizing layout area, 
minimizing the number of wire crossings in a layout, etc. (\cite{leighton1982}, \cite{leighton1984})

A realization of a bus graph conveys visibility relations among the buses and connectors. 
Given an arrangement of points (connectors) and axis-parallel line segments (buses) on a grid,
if a bus $B_i$  and a connector $c_j$ can be joined by a straight-line edge, then there exists 
an axis-parallel line of sight that does not intersect any buses or connectors
except $B_i$ and $c_j$. Furthermore, if all the buses are drawn horizontally, 
the bus graph $G$ is a subgraph of the visibility graph representing the vertical visibility
among the bus segments. There is an abundance of prior work on the visibility graphs based on
axis-parallel lines of sight, for example, in \cite{kaufmann96}, \cite{dean1997},
\cite{hutchinson1999}, \cite{whitesides2003}, \cite{wismath1985}, \cite{tamassia1986}, \cite{bose1996}, and~\cite{shermer1996}.
In particular, \cite{wismath1985} and \cite{tamassia1986} study bar visibility graphs (BVGs), 
where each vertex in the graph is drawn as a horizontal line segment in $\bbbr^2$, and the adjacency among
the vertices represent vertical visibility. It is shown that the recognition problem of 
such graphs can be solved in linear time. 
In \cite{dean1997}, \cite{whitesides2003}, \cite{hutchinson1999}, \cite{bose1996}, \cite{shermer1996},
the authors study rectangle visibility graphs (RVGs), where each vertex is drawn as a rectangle, 
and the adjacency among the vertices represent axis-parallel lines of sight. 
Reference \cite{shermer1996} shows that the recognition problem of such graphs is NP-complete.
The bus graphs that we study in this paper can be regarded as related to RVGs, where the vertices are 
restricted to degenerate rectangles such as line segments
and points. 

\section{Preliminaries}
Given a bus graph $G=(\mathcal{B}, \mathcal{C}; \mathcal{E})$, we call a vertex $B$ a $\mathcal{B}$-vertex
if $B\in \mathcal{B}$; its realization is called a \emph{bus}. Similarly, a $\mathcal{C}$-vertex refers to a vertex
in $\mathcal{C}$, and its realization is called a \emph{connector}. We often use
uppercase letters to denote $\mathcal{B}$-vertices, and lowercase letters to denote $\mathcal{C}$-vertices.

We use a function $\Gamma$ to denote an embedding of a combinatorial bus graph $G$. 
For example, $\Gamma(c)$ for some $\mathcal{C}$-vertex $c$ denotes the grid point where $c$ is laid out in
the embedding, and $\Gamma(B)$ for some $\mathcal{B}$-vertex $B$ or $\Gamma((B,c))$ for some edge $(B,c)$ 
denotes the line segment along a grid line where the bus or edge is laid out. 
We call a grid point an \emph{event point} if there is either a connector or an endpoint of a bus at that grid point.

It is important to note here that many variations of the bus graph problem can be devised, yet several are
equivalent. For example, suppose the buses are realized as \emph{open} line segments.
It is not hard to see that this variation is equivalent to the problem stated in Sect. 1, 
as an embedding with buses as open line segments can easily be transformed into an embedding with 
buses as closed line segments, and vice versa.

For another variation, note that a bus graph $G=(\mathcal{B}, \mathcal{C}; \mathcal{E})$ can be 
regarded as a hypergraph, where $\mathcal{B}$ is the set of vertices, and $\mathcal{C}$ is 
the set of hyperedges, each connecting at most four vertices. In this context, it is of interest 
to see if the realizability problem changes if we disallow multiple hyperedges in the hypergraph $G$. 
In other words, we would assume that no two $\mathcal{C}$-vertices are adjacent to the 
same set of $\mathcal{B}$-vertices. As the following lemma shows, this assumption does not change our problem.

\begin{lemma}
Let $G$ be a bus graph with multiple hyperedges in $\mathcal{C}$,
and let $G'$ be the bus graph constructed from $G$ by removing
hyperedge duplicates. Then $G$ is realizable if and only if $G'$ 
is realizable.
\end{lemma}
\begin{proof}
Necessity is trivial. Conversely, suppose $G'$ is realizable. 
Take any embedding of $G'$, and let $c$ be a connector in
the embedding at some position $(i, j)$. Then, create a copy of $c$ with the 
same connectivity to buses as follows. For each event point in the embedding,
if its $x$-coordinate is strictly greater than $i$, increase its
$x$-coordinate by 1. Then, if $c$ is connected to the rightmost endpoint of a horizontal bus $B$, 
where the endpoint is at an $x$-coordinate equal to $i$, increase the $x$-coordinate
of the endpoint by 1. Similarly, if an event point has a $y$-coordinate strictly
greater than $j$, increase its $y$-coordinate by 1. Then, if $c$ is connected to
the topmost endpoint of a vertical bus $B'$, where the endpoint is at a $y$-coordinate 
equal to $j$, increase the $y$-coordinate of the endpoint by 1. 
Finally, create a copy of $c$, and place it at $(i+1, j+1)$. 
Intuitively, we are \emph{stretching} the buses that intersect either of the two grid lines 
$x=i$ and $y=j$, so that there is a grid point for the newly created copy of $c$, without
making any collisions.
Repeat this for each $\mathcal{C}$-vertex in $G'$, until a realization of $G$ is obtained.
\qed
\end{proof}

\section{NP-Completeness}
\subsection{Membership in NP}
\begin{lemma}
Bus graph realizability is in NP.
\end{lemma}
\begin{proof} 
First, we claim that if a bus graph $G$ is realizable, there exists 
a compact layout such that the size of the layout is linear in 
each dimension. To see this, take any layout of $G$, and \emph{compact} the 
layout as follows. Take a vertical grid line with no event points on it.
Then, for each event point at $(x, y)$ appearing to the right of
this grid line, shift it to $(x-1, y)$. Observe that this operation
still guarantees a legal layout. Repeat this for each such vertical grid line.
Notice that the width of the layout is now linear in the size of $G$. 
We can apply a similar operation for all such horizontal grid lines. 
Thus, the claim is true, so there exists a short certificate for each 
realizable graph $G$ that gives the coordinates of the event points in a compact 
layout. It is easy to see that we can check, in polynomial time, if such a certificate 
represents a correct solution.\qed
\end{proof}

\subsection{NP-Hardness}
In this section, we prove the NP-hardness of BGR by a reduction from NAE-3SAT~\cite{garey1979}.
We first introduce and discuss properties of several gadgets, and then give the transformation.
\begin{definition}\label{def:ab-perp} An $(A, B)$-perp is a bus graph component consisting of three $\mathcal{C}$-vertices $x$, $y$, $z$, five $\mathcal{B}$-vertices $A$, $A'$, $B$, $B'$, $C$, and twelve edges $(A, x)$, $(A', x)$, $(B, x)$, $(B', x)$, $(A, y)$, $(A', y)$, $(B, y)$, $(C, y)$, $(A, z)$, $(A', z)$, $(B', z)$, $(C, z)$.
\end{definition}

A combinatorial graph of an $(A, B)$-perp and an example embedding of it are shown in Fig.~\ref{fig:ab-perp}.
\begin{lemma}\label{lemma:ab-perp} In any embedding $\Gamma$ of an $(A, B)$-perp,
\begin{enumerate}
\item $\Gamma(B)$ and $\Gamma(B')$ are parallel,
\item $\Gamma(A)$ and $\Gamma(B)$ are perpendicular, and
\item $\Gamma(A)$ and $\Gamma(A')$ are parallel.
\end{enumerate}
\end{lemma}
\begin{proof}\mbox{}
\begin{enumerate}
\item Note that $y$ and $z$ are both adjacent to $A$, $A'$ and $C$. $\Gamma((B, y))$ and $\Gamma((B', z))$ are 
then parallel, so $\Gamma(B)$ and $\Gamma(B')$ are parallel.
\item By 1, $\Gamma(B)$ and $\Gamma(B')$ are parallel. So if $\Gamma(A)$ were parallel to $\Gamma(B)$, 
then $x$ would be adjacent to three parallel buses. This is a contradiction.
\item By 2, $\Gamma(A)$ and $\Gamma(B)$ are perpendicular. If $\Gamma(A')$ were perpendicular to $\Gamma(A)$, then
  $\Gamma(A')$ would be parallel to $\Gamma(B)$, making $x$ adjacent to three parallel buses. This is a contradiction.
\end{enumerate}\qed
\end{proof}

\begin{definition}\label{def:bo-flipper} A $(B, o)$-flipper is a bus graph component consisting of an $(A, B)$-perp, one additional $\mathcal{C}$-vertex $o$, and two additional edges $(B, o)$, $(B', o)$.
\end{definition}

\begin{figure}[tb]\begin{center}
\begin{minipage}[b]{.4\textwidth}\begin{center}
\epsfig{file=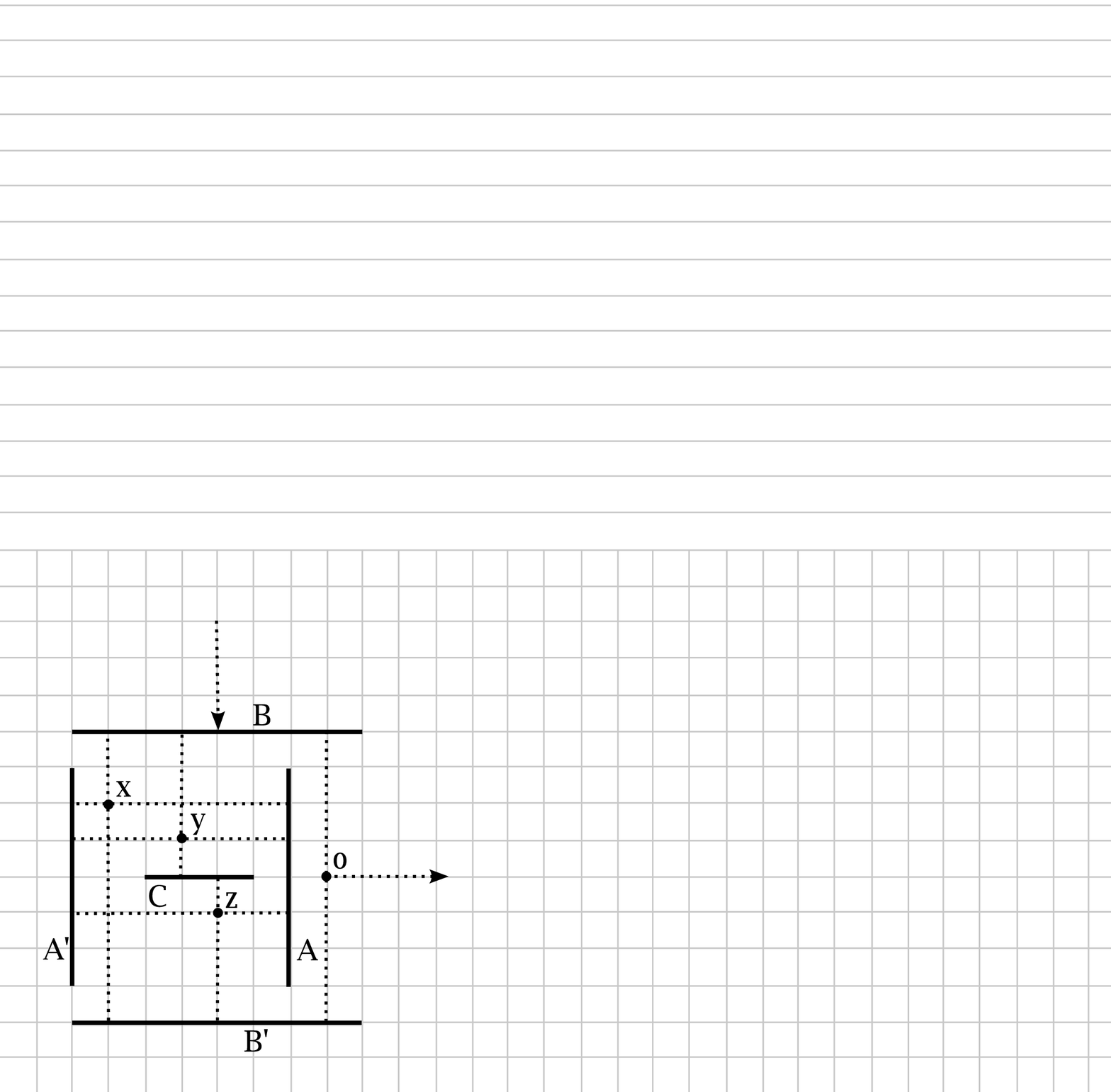, width=\textwidth, clip=}
\caption{Example embedding of a $(B, o)$-flipper.}\label{fig:bo-flipper}
\end{center}\end{minipage}
\qquad
\begin{minipage}[b]{.5\textwidth}\begin{center}
\epsfig{file=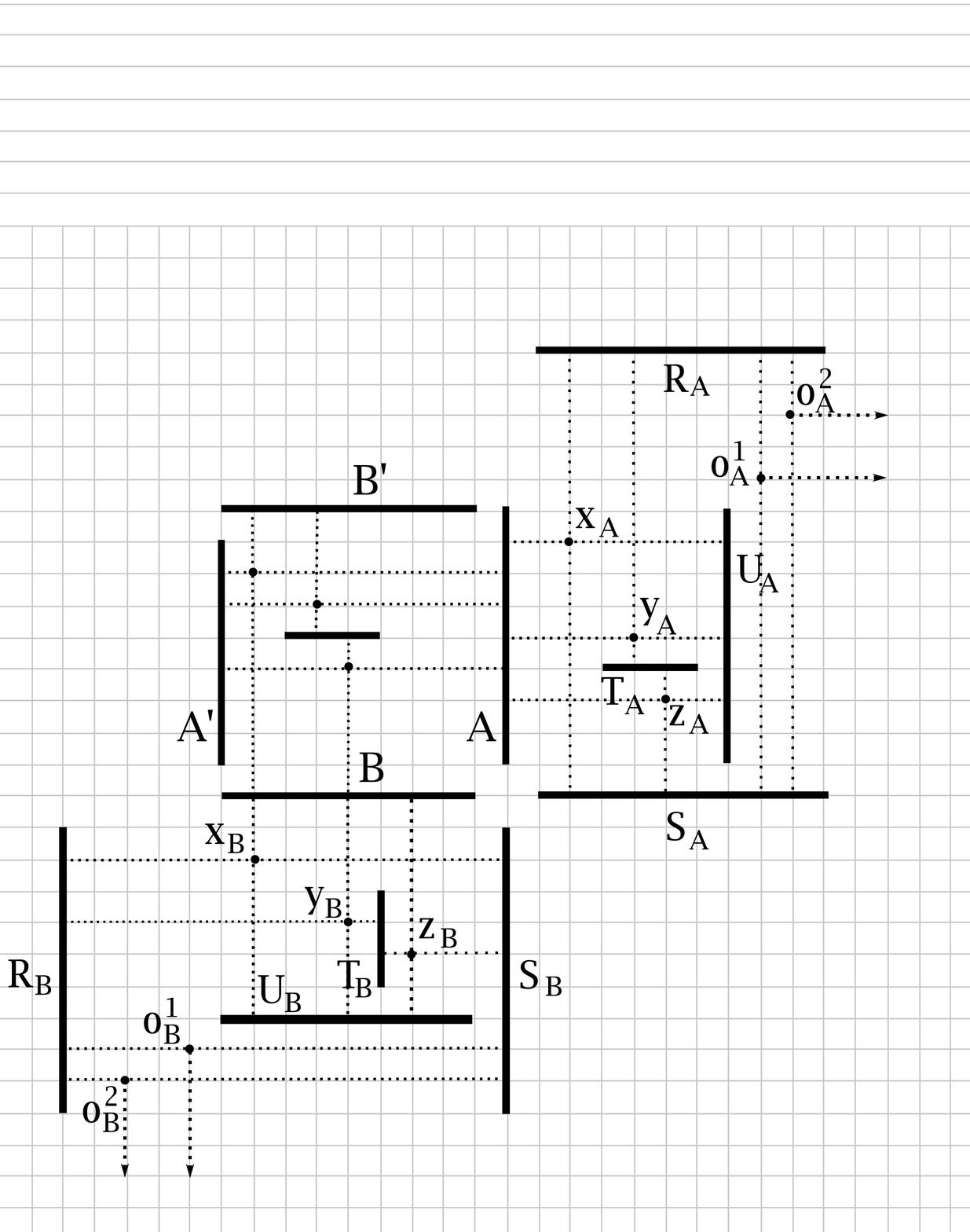, width=\textwidth, clip=}
\caption{Example embedding of an $(A, 2, B, 2)$-variable-box.}\label{fig:akbl-variablebox}
\end{center}\end{minipage}
\end{center}\end{figure}

An example embedding of a $(B, o)$-flipper is shown in Fig.~\ref{fig:bo-flipper}.

\begin{lemma}\label{lemma:bo-flipper} 
Let $i$ be a $\mathcal{C}$-vertex, and $O$ be a $\mathcal{B}$-vertex. If $i$ is joined with a $(B, o)$-flipper by
an edge $(B, i)$, and $O$ is joined with the $(B,o)$-flipper by an edge $(O, o)$, then in any embedding $\Gamma$,
\begin{enumerate}
\item $\Gamma((B, i))$ and $\Gamma((O, o))$ are perpendicular, and
\item $\Gamma(B)$ and $\Gamma(O)$ are perpendicular.
\end{enumerate}
\end{lemma}
\begin{proof}\mbox{}
\begin{enumerate}
\item By Lemma~\ref{lemma:ab-perp}, $\Gamma(B)$ and $\Gamma(B')$ are parallel, 
so $\Gamma((B, o))$ and $\Gamma((B', o))$ are also parallel. $\Gamma((O, o))$ is then 
perpendicular to $\Gamma((B, o))$. Thus, $\Gamma((O, o))$ is parallel to $\Gamma(B)$ and 
perpendicular to $\Gamma((B, i))$.
\item Follows immediately from 1.
\end{enumerate}\qed
\end{proof}

\begin{definition}\label{def:akbl-variablebox} An $(A, k, B, l)$-variable-box is a bus graph component consisting of an $(A,B)$-perp, $(3 + k) + (3 + l)$ additional $\mathcal{C}$-vertices $x_A$, $y_A$, $z_A$, $o^1_A$, $o^2_A$, \ldots, $o^k_A$, $x_B$, $y_B$, $z_B$, $o^1_B$, $o^2_B$, \ldots, $o^l_B$, eight additional $\mathcal{B}$-vertices $R_A$, $S_A$, $T_A$, $U_A$, $R_B$, $S_B$, $T_B$, $U_B$, and $(12 + 2k) + (12 + 2l)$ additional edges
\begin{eqnarray*}
(A, x_A), (R_A, x_A), (S_A, x_A), (U_A, x_A),&\\
(A, y_A), (R_A, y_A), (T_A, y_A), (U_A, y_A),&\\
(A, z_A), (S_A, z_A), (T_A, z_A), (U_A, z_A),&\\
(R_A, o^i_A), (S_A, o^i_A) &\mbox{ for $i = 1, 2, \ldots, k$,}\\
(B, x_B), (R_B, x_B), (S_B, x_B), (U_B, x_B),&\\
(B, y_B), (R_B, y_B), (T_B, y_B), (U_B, y_B),&\\
(B, z_B), (S_B, z_B), (T_B, z_B), (U_B, z_B),&\\
(R_B, o^i_B), (S_B, o^i_B) &\mbox{ for $i = 1, 2, \ldots, l$.}
\end{eqnarray*}
\end{definition}

An example embedding of an $(A, 2, B, 2)$-variable-box is shown in Fig.~\ref{fig:akbl-variablebox}.

\begin{lemma}\label{lemma:variablebox} Let $(O^i_A, o^i_A)$ for $i = 1, 2, \ldots, k$ and $(O^j_B, o^j_B)$ for $j = 1, 2, \ldots, l$ 
be edges joined with an $(A, k, B, l)$-variable-box. Then in any embedding $\Gamma$,
\begin{enumerate}
\item $\Gamma((O^i_A, o^i_A))$ and $\Gamma(A)$ are perpendicular for any $i = 1, 2, \ldots, k$.
\item $\Gamma((O^i_B, o^i_B))$ and $\Gamma(B)$ are perpendicular for any $i = 1, 2, \ldots, l$.
\end{enumerate}
\end{lemma}
\begin{proof} We prove the first statement; the proof of the second is analogous. Notice that vertices $x_A$, $y_A$, $z_A$, $R_A$, $S_A$, $T_A$, $U_A$, $o^1_A$, $o^2_A$, \ldots, $o^k_A$ form an $(R_A, o^i_A)$-flipper with multiple output $\mathcal{C}$-vertices. 
As shown in the proof of Lemma~\ref{lemma:bo-flipper}, $\Gamma((O^i_A, o^i_A))$ is parallel to $\Gamma(R_A)$ for all $i=1, 2,\ldots, k$. Using the same reasoning as in the proof of Lemma~\ref{lemma:ab-perp}, $\Gamma(R_A)$ is perpendicular to $\Gamma(A)$. It follows that  
every edge $\Gamma((O^i_A, o^i_A))$ is perpendicular to $\Gamma(A)$.\qed\end{proof}

\begin{definition}\label{def:io-chain} An $(I, O)$-chain is a bus graph component consisting of
\begin{enumerate}
\item an $(I, o_1)$-flipper,
\item an $(I_1, o_2)$-flipper,
\item an $(I_2, o_3)$-flipper,
\item an $(I_3, O)$-perp,
\end{enumerate}
and three additional edges $(I_1, o_1)$, $(I_2, o_2)$, $(I_3, o_3)$.
\end{definition}

\begin{figure}[tb]
\begin{center}
\epsfig{file=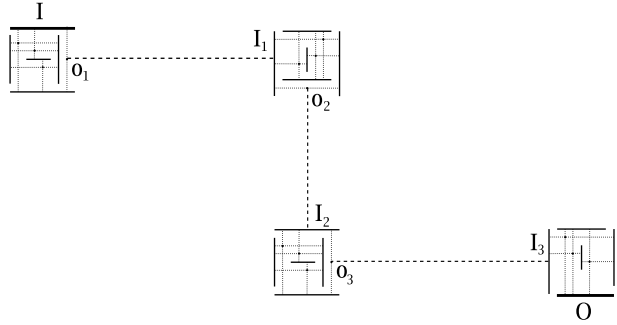, height=5cm, clip=}
\end{center}
\caption{Example embedding of an $(I, O)$-chain.}
\label{fig:io-chain}
\end{figure}

An example embedding of an $(I, O)$-chain is shown in Fig.~\ref{fig:io-chain}.

\begin{lemma}\label{lemma:io-chain} In any embedding $\Gamma$ of an $(I, O)$-chain, $\Gamma(I)$ and $\Gamma(O)$ are parallel.\end{lemma}
\begin{proof} By repeatedly applying Lemma~\ref{lemma:bo-flipper}, $\Gamma(I)$ is perpendicular to $\Gamma(I_1)$, $\Gamma(I_1)$ is perpendicular to $\Gamma(I_2)$, and $\Gamma(I_2)$ is perpendicular to $\Gamma(I_3)$. By Lemma~\ref{lemma:ab-perp}, $\Gamma(I_3)$ is perpendicular to $\Gamma(O)$. It follows that $\Gamma(I)$ is parallel to $\Gamma(O)$.\qed\end{proof}

Finally, we are ready to give the transformation from NAE-3SAT to BGR. 
Let $\phi$ be an instance of NAE-3SAT, consisting of boolean variables $x_1$, \ldots, $x_n$, 
and clauses $C_1$, \ldots, $C_m$. Construct a bus graph $G$ as follows.

\begin{enumerate}
\item For each boolean variable $x_i$, create a $(X_i, t_i, \overline{X}_i, f_i)$-variable-box, where $t_i$ and $f_i$ are the numbers of distinct occurrences of the literals $x_i$ and $\overline{x_i}$, respectively, in $\phi$.
\item For each clause $C_q = (x^*_i \lor x^*_j \lor x^*_k)$, where $x^*$ is either $x$ or $\overline{x}$, create
\begin{enumerate}
\item a $\mathcal{C}$-vertex $c_q$,
\item an $(I_{q,1}, O_{q,1})$-chain, an $(I_{q,2}, O_{q,2})$-chain, and an $(I_{q,3}, O_{q,3})$-chain,
\item edges $(O_{q,1}, c_q)$, $(O_{q,2}, c_q)$ and $(O_{q,3}, c_q)$,
\item edges $(I_{q,1}, p_i)$, $(I_{q,2}, p_j)$ and $(I_{q,1}, p_k)$,
  where $p_i = o^r_{X_i}$ if $x^*_i = x_i$ and $p_i = o^r_{\overline{X}_i}$ 
if $x^*_i = \overline{x_i}$, and it is the $r$th occurrence of $x^*_i$ being considered.
\end{enumerate}
\end{enumerate}

Since every gadget is of linear size, the transformation clearly takes polynomial time. Finally, the following lemma 
completes the proof of Theorem~\ref{thm1}.

\begin{figure}[tb]
\begin{center}
\epsfig{file=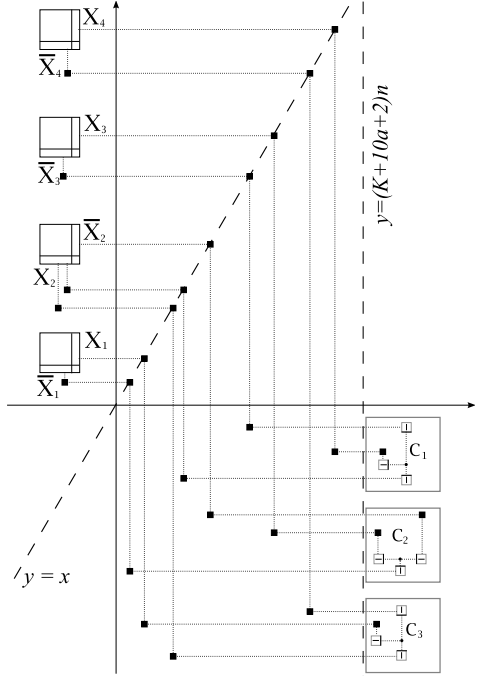, height=9cm, clip=}
\end{center}
\caption{A schematic embedding of $G$, where $\phi$ consists of the clauses 
$C_1=(x_2 \vee \overline{x_3} \vee x_4)$,
$C_2=(\overline{x_1} \vee \overline{x_2} \vee x_3)$,and
$C_3=(x_1 \vee x_2 \vee \overline{x_4})$ with a satisfying truth 
assignment $x_1 = x_3 = x_4 =$ true and $x_2=$false. Note the regions
separated by the dashed lines and $x-, y-$axes. Complete instructions
for the embedding is included in Appendix.
}
\label{fig:yes-to-yes}
\end{figure}

\begin{lemma}\label{lemma:reduction} 
$\phi \in$ NAE-3SAT if and only if $G \in BGR$.
\end{lemma}
 
\begin{proof} 
Suppose $\phi \in $ NAE-3SAT. An embedding $\Gamma$ of $G$ can be constructed as 
demonstrated in Fig.~\ref{fig:yes-to-yes}. Notice that the 
variable boxes are embedded in such a way that the buses corresponding to true literals are 
drawn vertically and the buses corresponding to false literals are drawn horizontally. 
See Appendix for a full description of this layout.

Conversely, suppose $G \in $ BGR, and take an embedding $\Gamma$ of $G$. 
By Lemma~\ref{lemma:ab-perp}, $\Gamma(X_i)$ is perpendicular to $\Gamma(\overline{X}_i)$
for each variable box, so assign each variable $x_i$ to be true if $\Gamma(X_i)$ 
is vertical and false otherwise. To see that this truth-assignment satisfies the clauses, 
consider a clause $C_q = (x^*_i \lor x^*_j \lor x^*_k)$. The clause vertex $c_q$ is adjacent
to three buses $O_{q, 1}$, $O_{q,2}$, and $O_{q,3}$ at the end of $(I, O)$-chains. Since $c_q$ can be joined to at most 
two parallel buses, at least one of these three buses must be drawn horizontally, and at 
least one must be drawn vertically. Take any one of the three buses, say $O_{q,1}$, and 
consider the literal bus to which $O_{q,1}$ connects in the variable box. By 
Lemma~\ref{lemma:variablebox} and Lemma~\ref{lemma:io-chain}, the orientation of these two 
buses must be the same. This implies that the clause vertex $c_q$ is connected to at least 
one vertically drawn literal bus, and at least one horizontally drawn literal bus. Therefore, the
truth-assignment satisfies $\phi$.
\qed

\end{proof}

\section{Applications of Proof Techniques}
In this section, we look at variations of the bus graph realizability problem
in which the degree of $\mathcal{C}$-vertices is restricted. In the 
original bus graph realizability problem presented in Sect. 1, 
each $\mathcal{C}$-vertex has a maximum degree of 4, due to the orthogonal drawing style. 
An analogous problem can be devised for the class of bus graphs where the $\mathcal{C}$-vertices 
have maximum degrees of either 2 or 3. In what follows, we show that these variations are also NP-complete.
Finally, we discuss the problem of merely deciding the orientation of buses, which also turns out to 
be NP-complete.

\subsection{Connectors with Bounded Degree}
First, consider the class of bus graphs where the $\mathcal{C}$-vertices have maximum degree 1. 
These graphs are trivially realizable by simply drawing all the buses along a grid line.
However, if the maximum degree of $\mathcal{C}$-vertices is greater than or equal to 2, the problem
becomes harder. Recall from the proof of Theorem~\ref{thm1} in Sect. 3 that the reduction from NAE-3SAT
was done using a series of gadgets, each of which was based on the $(A, B)$-perp. The following results 
follow from constructing an $(A, B)$-perp gadget for each case, and then
constructing the remaining gadgets in a similar fashion. 

\begin{theorem}
  \label{thm:degree2}
  Bus Graph Realizability is NP-complete when the maximum degree of $\mathcal{C}$-vertices is
  2.
\end{theorem}

\paragraph{Proof sketch.}
If every $\mathcal{C}$-vertex has degree 2, we may regard each $\mathcal{C}$-vertex as
an edge connecting two $\mathcal{B}$-vertices. Thus, in the following description, when we
say to join two $\mathcal{B}$-vertices, we mean to create a $\mathcal{C}$-vertex that joins the
two $\mathcal{B}$-vertices. To construct the perp gadget, start with a complete graph $K_8$, with vertices labeled 
$B_1$ through $B_8$. Then, create another vertex $B_9$, and join it with vertices $B_1$ 
through $B_7$. Finally, create a vertex $B_{10}$, and join it to vertices $B_1$ 
through $B_6$, $B_8$, and $B_9$. It can be shown that the buses $B_8$ and $B_{10}$ must be
drawn perpendicularly to each other, and all other gadgets can be constructed based on 
this structure.\footnote{
  For obvious reasons, we cannot use a single $\mathcal{C}$-vertex to represent a clause. 
  This can be resolved by creating a combination of perp gadgets and $(I, O)$-chains.
}
\qed

\begin{theorem}
  \label{thm:degree3}
  Bus Graph Realizability is NP-complete, when the maximum degree of $\mathcal{C}$-vertices is
  3.
\end{theorem}

\paragraph{Proof sketch.}
The perp gadget for this case is a bus graph component consisting of eight $\mathcal{C}$-vertices 
$s$, $t$, $u$, $v$, $w$, $x$, $y$, $z$, six $\mathcal{B}$-vertices $A$, $A'$, $B$, $B'$, $C$, $D$ 
and 24 edges, connecting the $\mathcal{B}$-vertices to the $\mathcal{C}$-vertices as follows: 
$s$ to $A$, $B$, $C$; $t$ to $A$, $B$, $A'$; $u$ to $A$, $B$ ,$B'$; $v$ to $C$, $A'$, $B'$; 
$w$ to $A'$, $B$, $D$; $x$ to $A'$, $B$, $B'$; $y$ to $D$, $A$, $B'$; $z$ to $A'$, $B'$, $A$. 
It can be shown that the buses $A$ and $B$ must be drawn perpendicularly to each other,
and parallel to $A'$ and $B'$, respectively. All other gadgets can be constructed 
based on this structure. \qed

Our results can be summarized as follows.

\begin{theorem}
  Bus Graph Realizability is NP-complete if and only if the maximum degree of $\mathcal{C}$-vertices 
  in the given graph is 2, 3, or 4.
\end{theorem}

\subsection{Partition by Orientation}

In order to realize a given bus graph, one must decide the 
orientations of the buses. Since a connector can be joined to at most
two horizontal buses and at most two vertical buses, all realizable 
bus graphs admit a bipartition of buses by orientation. As we will
see shortly, even deciding whether the buses can be properly oriented
is hard to compute. Note, however, that a proper bipartition by orientation
does not guarantee that the graph is realizable.\footnote{
  A simple counterexample can be constructed by first creating a complete 
  graph with 9 $\mathcal{B}$-vertices, and then placing a $\mathcal{C}$-vertex 
  on each edge. 
}

We define the problem \emph{PARTITION-BY-ORIENTATION} as follows:

\begin{description}
\item \textbf{Instance}: A bipartite graph $G=(\mathcal{B}, \mathcal{C}; \mathcal{E})$ 
such that $\forall c\in \mathcal{C}$, $deg(c)\leq 4$.
\item \textbf{Question}:  Can $\mathcal{B}$ be partitioned into two disjoint sets $\mathcal{B_H}$ and $\mathcal{B_V}$, 
    such that $\forall c\in \mathcal{C}$, $c$ is adjacent to no more than two vertices in $\mathcal{B_H}$
    and no more than two vertices in $\mathcal{B_V}$?
\end{description}

\begin{theorem}
\label{thm:orientation}
  PARTITION-BY-ORIENTATION is NP-complete.
\end{theorem}

\begin{proof}

The problem is clearly in NP, as one could guess a partition and check for its
correctness in polynomial time. For the reduction, we use NAE-3SAT. 
Given an instance $\phi$ of NAE-3SAT, we construct a graph $G$ as follows. 
First, for each variable $x_i$ in $\phi$, create an $(A, B)$-perp. 
Vertex $A$ corresponds to $x_i$, whereas vertex $B$ corresponds to $\overline{x_i}$. 
Then, for each clause $C_q=(x^*_i \vee x^*_j \vee x^*_k)$, create a $C$-vertex $c_q$, 
and join it to the three literal vertices. The rest of the proof is similar to the proof of 
Lemma~\ref{lemma:reduction}, and follows from Lemma~\ref{lemma:ab-perp}. \qed

\end{proof}

\section{Bus Graph Realizability with Given Bus Lengths}
In this section, we study a variation of bus graph realizability in which
the lengths of buses are given as input (BGR+BL). We devise a new encoding scheme for
the instances of this problem. The purpose of this encoding scheme will become
clear shortly. Recall from the discussion in Sect. 2 that bus graphs can be
regarded as hypergraphs, where each $\mathcal{C}$-vertex corresponds to a hyperedge.
In the new encoding scheme, we assume that the problem instance is given
as a list of subsets of $\mathcal{B}$, where each subset is of cardinality at most 4. 
For each subset, we also encode the number of $\mathcal{C}$-vertices that are adjacent to exactly that subset 
of $\mathcal{B}$-vertices. Notice that this is a form of adjacency matrix for
the hypergraph, where each entry in the matrix denotes the number of hyperedges
for that particular subset of buses. In the case of hypergraphs where no multiple
hyperedges are allowed, each entry in the matrix would be either 1 or 0. 

Now we are ready to state and prove the main result of this section.

\begin{theorem}
BGR+BL is NP-hard. It is also NP-hard if the maximum degree of $\mathcal{C}$-vertices
is 2, or if we require the buses to be parallel to each other.
\end{theorem}
\begin{proof}
The reduction is from PARTITION~\cite{garey1979}. Let $\langle A, s\rangle$ be an instance of PARTITION,
where $A$ is a set of elements, and $s:A\rightarrow Z^+$ is a size function for each
element. For simplicity, we assume that no element is of size 1, as we can scale 
the size function appropriately. We construct a bus graph $G$ as follows. 
Create a $\mathcal{B}$-vertex $B^*$ of length $\frac{1}{2} \sum_{a \in A} s(a) - 1$. 
Then, for each $a \in A$, create a $\mathcal{B}$-vertex $B_a$ of length $s(a)-1$. 
Now, for each element $a\in A$, create exactly $s(a)$ copies of
$\mathcal{C}$-vertex, and join them to both $B_a$ and $B^*$. With the new encoding
scheme, this transformation can be done in polynomial time. 

Suppose $\langle A, s\rangle\in$ PARTITION. First, lay out the bus $B^*$
horizontally along the $x$-axis. It is easy to see that one can 
place the buses $B_a$ for all $a\in A'$ horizontally along the grid line $y=1$ (above $B^*$),
and place the rest horizontally along the grid line $y=-1$ (below $B^*$). This is a legal layout.

Conversely, suppose $G$ is a yes-instance of BGR+BL. Take an embedding of $G$, and assume without
loss of generality that $B^*$ is drawn horizontally. Then, each bus $B_a$ must be laid out either
completely above $B^*$, or completely below $B^*$. To see this, suppose otherwise.
This means either (1) some $B_a$ is drawn on the same grid line as $B^*$, or (2) 
some $B_a$ is drawn vertically, where one endpoint of $B_a$ is above $B^*$, and the other
endpoint is below $B^*$. Case (1) is not possible, as no connector can join $B_a$
with $B^*$. Consider case (2).
Then, there exists a grid point $p$ at the intersection
of $B_a$ and the grid line where $B^*$ is drawn. No connector can connect $p$ with $B^*$,
since every edge must connect to a bus ($B^*$, in particular) perpendicularly. This is a contradiction because
each grid point along the bus $B_a$ must connect to  $B^*$ via a connector.

This implies that the number of grid points along the buses drawn above $B^*$ equals
the number of grid points along $B^*$. Similarly, the number of grid points along the buses
drawn below $B^*$ equals the number of grid points along $B^*$.
Therefore, we have a subset $A'\subset A$, where $A'$ is the set of elements
that correspond to the buses drawn above $B^*$, and this is a valid partition.
\qed
\end{proof}


\section{Concluding Remarks and Open Problems}

Although bus graph realizability is an NP-complete problem in general,
some special classes of graphs admit polynomial time solutions. For example, 
if the given bus graph $G$ is a tree, it is simple to devise an algorithm to produce a 
realization of $G$, and hence $G$ \emph{always} admits a bus graph embedding. 
However, what other classes of graphs admit polynomial time recognition 
algorithms for realizable bus graphs is an unexplored question.

As a consequence of the hardness results of this paper, one may search for
approximate solutions to the problems. It is unclear, however, what
optimization criteria would be used. With applications in VLSI in mind, one may
wish to lay out all the buses first, and then maximize the connectivity by maximizing
the number of connectors realized in the layout. 

\subsubsection{Acknowledgments.}
We thank Olivier Mireault for his interest and support. Those of us who hold research grants or 
government scholarships gratefully acknowledge NSERC and FQRNT for their support. 

\bibliographystyle{splncs} 
\bibliography{busgraph}

\pagebreak
\section*{Appendix: Necessity Proof for Lemma~\ref{lemma:reduction}}

In this section, a full proof for the necessity of the Lemma~\ref{lemma:reduction} is given.
Let $\phi$ denote a boolean formula for NAE-3SAT, consisting of
$n$ variables and $m$ clauses, and let $\sigma:X\rightarrow\{T, F\}$ be a satisfying 
truth assignment for $\phi$. Given $\phi$, our goal is to show that the bus graph $G$, 
constructed from the transformation shown in Sect. 3.2, is realizable. 
We first describe how each gadget, as introduced in Sect. 3.1, can be embedded
within a bounding box of predefined size. Then, the overall embedding of 
$G$ is presented by explicitly giving coordinates for each gadget bounding box. 
Finally, we show that the so-described embedding of $G$ is a legal layout of a bus graph, which completes the proof.
For simplicity, we often refer to a bounding box as an $n\times n$ \emph{subgrid}, which is a square of
size $(n-1)\times (n-1)$, with $n$ grid points along each side. 

\subsection*{A. Embedding of Gadgets}
Note that the coordinates used in this section are relative to each corresponding gadget only.  

\subsubsection{Perps and Flippers.} 
We embed a perp within an $9\times 9$ subgrid, as shown in Fig.~\ref{fig:ab-perp}. 
In the case of the $(I_3, O)$-perp inside an $(I, O)$-chain, its incoming
edge is connected to the grid point at the center of the bus $\Gamma(I_3)$. See Fig.~\ref{fig:io-chain} for an
example. Similarly, a flipper is embedded in a $9\times 9$ subgrid, and the
incoming and outgoing edges are drawn so that they lie along the same grid line as the
center grid point of the subgrid. See Fig.~\ref{fig:bo-flipper} for an illustration of the embedding.

\begin{figure}[tbh]
\begin{center}
\epsfig{file=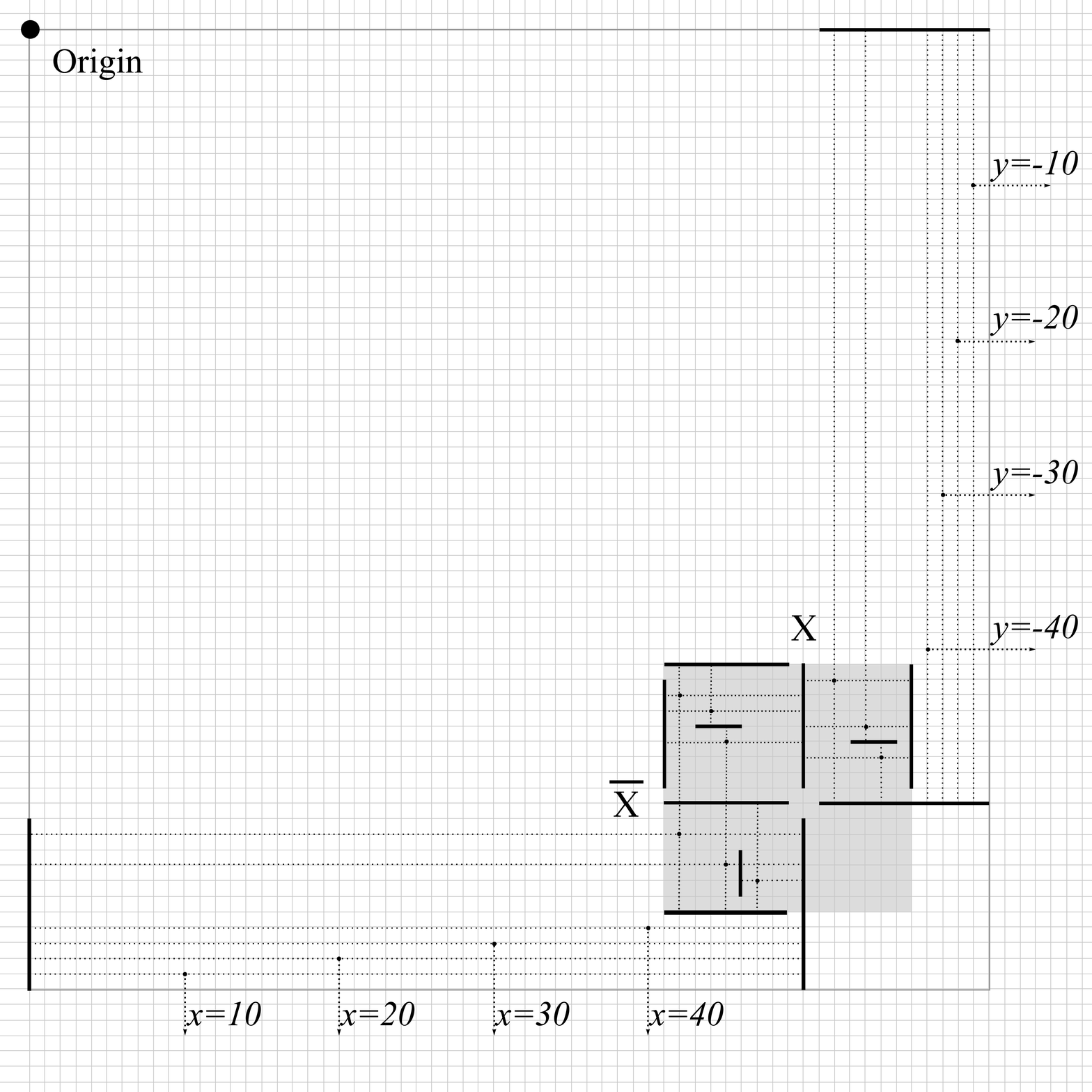, height=7cm, clip=}
\end{center}
\caption{
Embedding of a $(X,4,\overline{X},4)$-variable box. 
Observe that there are 16 grid points along each side of the shaded region, and
$(10a +1)+(a + 1)$ grid points on either side of the shaded region. The orientation
in the figure assumes $X=$true.
}

\label{fig:variablebox} 
\end{figure}

\subsubsection{Variable Boxes.}
Let $a$ denote the maximum number of occurrences of a literal in $\phi$.
Then, draw each variable box in a $K\times K$ subgrid, where $K=11a + 18$. 
Then, using the upper-left corner of the subgrid as the origin:
\begin{enumerate}
  \item Embed the horizontal outgoing edges along the grid lines $y=-10i$
  \item Embed the vertical outgoing edges along the grid lines $x=10i$
\end{enumerate}
for all $i=1\ldots a$. See Fig.~\ref{fig:variablebox} for the embedding of a $(X, 4, \overline{X}, 4)$-variable box.
Notice that some literals may not appear $a$ times in $\phi$, but
the embedding reserves a fixed area for every variable box to accommodate 
all the occurrences of a literal in $\phi$. 
Notice that all parallel outgoing edges are spaced apart by 9 grid lines.
It is also important to note that the embedding is
symmetric along the diagonal of the subgrid with respect to the position of
outgoing edges. 
This is crucial for the embedding, as the orientation of the embedding is decided
upon the satisfying assignment $\sigma$. If a variable $X_i$ is assigned true 
by $\sigma$, then the corresponding variable box is embedded so that
the bus $\Gamma(X_i)$ is drawn vertically. Otherwise, it is embedded as a reflection
about the diagonal so that $\Gamma(X_i)$ is drawn horizontally.

\subsubsection{Clause Boxes.}
\begin{figure}[tbh]
\begin{center}
\epsfig{file=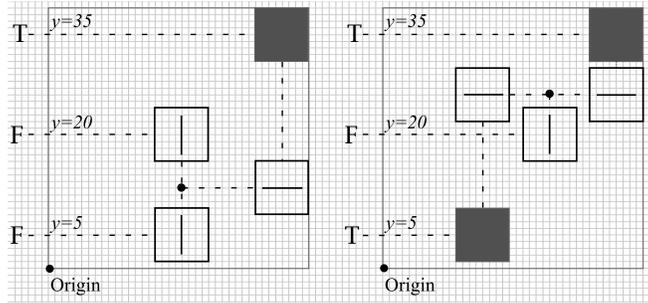, height=4cm, clip=}
\end{center}
\caption{
Embedding for a clause box assigned with $T,F,F$ (\emph{left}), and 
a clause box assigned with $T,T,F$ (\emph{right}). Shaded boxes
denote flipper components, whereas unshaded boxes denote perp components,
for each corresponding $(I, O)$-chain.
}
\label{fig:clausebox}
\end{figure}

Although not presented as a gadget in Sect. 3.2, each clause connector $c_i$ can be embedded 
on a $40\times 40$ subgrid, along with some components of $(I,O)$-chains joined to $c_i$. 
As with the embedding of variable boxes, the embedding of clause boxes depends on 
$\sigma$. Since $\sigma$ is a satisfying assignment of NAE-3SAT, there are 2 cases for
the truth assignment of each clause: $(T, F, F)$, and $(T, T, F)$. See Fig.~\ref{fig:clausebox}
for the embedding of both cases. Observe that if a literal in the clause is assigned to true,
then the $(I_2, o_3)$-flipper and $(I_3, O)$-perp component in the corresponding $(I,O)$-chain 
is drawn inside the subgrid. Otherwise, only the $(I_3, O)$-perp component is drawn
inside the subgrid. Observe that the incoming edges are drawn at fixed $y$-coordinates, 
regardless of the assignment for the clause. 

\subsection*{B. Embedding of $G$}
 
We are now ready to realize $G$. For an overview of the embedding, 
see Fig.~\ref{fig:yes-to-yes}. First, we begin by embedding the variable boxes. 
For each boolean variable $X_i$, lay out the corresponding variable box on the subgrid
bounded by $x\in [-K, -1]$ and $y\in [(K + 10a + 2)i - K + 1, (K + 10a + 2)i]$, in
the orientation determined by $\sigma$. 
Observe that each variable box is embedded in the second quadrant, and every 
two neighboring subgrids are $10a + 2$ grid lines apart from each other.

Then, for each clause $c_q$, lay out the corresponding clause box on the subgrid 
bounded by $x\in[(K+10a+2)n,(K+10a+2)n+39]$ and $y\in[-40q,-40q+39]$. The internal
layout of each clause box is determined by $\sigma$, but the three incoming edges
are always located at $y=-40q + 5, y=-40q + 20$, and $y=-40q + 35$.
Observe that each clause box is embedded to the right of the grid line $y=(K+10a+2)n$.

\begin{figure}[tbh]
\begin{center}
\epsfig{file=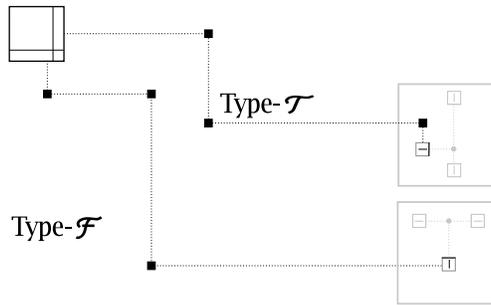, height=4cm, clip=} 
\end{center}
\caption{
Type-$\mathcal{T}$ embedding and Type-$\mathcal{F}$ embedding
}
\label{fig:tfchains}
\end{figure}

We now need to lay out the $(I, O)$-chains that join the variable boxes with clause boxes. 
There are two types of embedding for $(I, O)$-chains, determined by the truth-assignment 
of the literal bus that the $(I, O)$-chain connects to.
If an $(I, O)$-chain connects to a literal assigned true, we say that its 
embedding is of Type-$\mathcal{T}$.
If an $(I, O)$-chain connects to a literal assigned false, we say that its 
embedding is of Type-$\mathcal{F}$. 
We describe the two embedding types separately. See Fig.~\ref{fig:tfchains} for an illustration.
  
\subsubsection{Type-$\mathcal{T}$ Embedding.}
Let $c_q$ be the clause vertex inside the clause box, and let $o^{r_i}_{X_i}$ be the literal vertex
inside the variable box. Our aim is to connect $c_q$ to $o^{r_i}_{X_i}$ using an $(I, O)$-chain.
For Type-$\mathcal{T}$, recall that $(I_2, o_3)$-flipper and $(I_3, O)$-perp are embedded within
the clause box. Thus, it suffices to describe the positions for the $(I, o_1)$-flipper and 
the $(I_1, o_2)$-flipper.

Observe that the $(I, O)$-chain must exit the variable box horizontally at $y=y_1$, where $y_1$ is the $y$-coordinate
of $o^{r_i}_{X_i}$, and enter the clause box horizontally at $y=y_2$, where $y_2$ is the position defined 
by the clause box type as shown in Fig.~\ref{fig:clausebox}.
Place the two flippers as follows.

\begin{enumerate}
\item Embed the $(I, o_1)$-flipper in the subgrid bounded by $x, y\in [y_1 -4, y_1 + 4]$.
\item Embed the $(I_1, o_2)$-flipper in the subgrid bounded by $x\in [y_1 -4, y_1 + 4]$ and $y\in [y_2 - 4, y_2 + 4]$.
\end{enumerate}

\subsubsection{Type-$\mathcal{F}$ Embedding.}
Let $c_q$ be the clause vertex inside the clause box, and let $o^{r_j}_{\overline{X}_j}$ be the literal vertex
inside the variable box. Our aim is to connect $c_q$ to $o^{r_j}_{\overline{X}_j}$ using an $(I, O)$-chain.
For Type-$\mathcal{F}$, only the $(I_3, O)$-perp is embedded within the clause box. Hence, we must
describe the positions for $(I, o_1)$-flipper, $(I_1, o_2)$-flipper, and $(I_2, o_3)$-flipper. 

Observe that the $(I, O)$-chain must exit the variable box vertically at $x=x_1$, where $x_1$ is the $x$-coordinate
of $o^{r_j}_{\overline{X}_j}$, and enter the clause box horizontally at $y=y_2$, where $y_2$ is the position defined 
by the clause box type as shown in Fig.~\ref{fig:clausebox}.
Place the three flippers as follows.

\begin{enumerate}
\item Embed the $(I, o_1)$-flipper in the subgrid bounded by $x\in[x_{1}-4,x_{1}+4]$ and 
$y\in[y_1 -4, y_1 + 4]$, where $y_1=(K+10a+2)(j-1)+6+10(r_{j}-1)$.
\item Embed the $(I_1, o_2)$-flipper in the subgrid bounded by $x,y\in[y_{1}-4,y_{1}+4]$. 
\item Embed the $(I_2, o_3)$-flipper in the subgrid bounded by $x\in[y_{1}-4,y_{1}+4]$ and $y\in[y_{2}-4,y_{2}+4]$.
\end{enumerate}

\subsection*{C. Correctness of the Embedding}
In this section, we show that the embedding of $G$ is a legal layout.
To do this, we refer back to the four properties stated in the problem definition in Sect. 1. 
The first two properties hold trivially by construction. Hence, it suffices to show that the other two properties 
hold.

\begin{proof}

We check if the properties hold as we lay out each gadget.
By construction, the embedding of each individual gadget is legal. 
So first lay out the variable boxes and clause boxes. 
The embedding of these boxes together is legal, since no two boxes overlap.
Now, lay out the $(I, O)$-chains and see if the two
properties still hold. 

\noindent(Property 3.) We say an edge \emph{passes through} a gadget if the embedded edge
enters and exits the bounding box of the gadget.
Since each bus or connector is drawn within the bounding box of a gadget,
we need to show that no edge passes through a gadget. Note that we need only
consider the edges along the $(I, O)$-chains, as all other edges are embedded inside gadget bounding
boxes. Notice that for each flipper $f$ in some $(I, O)$-chain in the embedding, 
$f$ is joined to other components by a horizontal edge $f_H$ and a vertical edge $f_V$. 

Now, take a horizontal edge $e_H$ on some grid line $y=y_1$. There are three cases to consider.
(1)  Suppose $e_H$ passes through a flipper $f$ in some $(I,O)$-chain. 
Then $e_H$ must be fewer than 4 grid lines apart from $f_H$. This is a contradiction, 
as any two horizontal edges in the embedding are at least 9 grid lines apart from each other. 
(2)  Suppose $e_H$ passes through a variable box. If $y_1>0$, there is exactly one 
  variable box at $y=y_1$, which $e_H$ connects to at its endpoint. If $y_1\leq0$, $e_H$ cannot
  reach any variable box because all variable boxes are above the $x$-axis. 
  This is a contradiction.
(3)  Suppose $e_H$ passes through a clause box. If $y_1\leq 0$, there is exactly 
  one clause box at $y=y_1$, which $e_H$ connects to at its endpoint. 
  If $y_1>0$, $e_H$ cannot reach any clause box because all clause boxes are below the $y$-axis,
  hence a contradiction.

Take a vertical edge $e_V$ on some grid line $x=x_1$. Note that $e_V$ cannot pass 
through a clause box, since all vertical edges lie to the left of the grid line $x=(K+10a+2)n$. 
Secondly, $e_V$ cannot pass through a variable box,
since the $y$-axis separates $e_v$ from variable boxes when $x_1>0$, or
$e_v$ completely lies within region between two neighboring variable boxes when $x_1 \leq 0$.
Finally, $e_v$ cannot pass through flippers, either because the vertical
edge $f_V$ for any flipper $f$ is sufficiently far from $e_v$, or $e_v$ and $f$ 
lie in different regions separated by variable boxes.

\noindent(Property 4.) We say two gadgets \emph{intersect} if the bounding boxes of
the two embedded gadgets intersect. 
Since each bus or connector is embedded within some gadget, 
if no two gadgets intersect then no buses or connectors 
may intersect. We categorize the gadgets into 5 types as follows.
\begin{enumerate}
  \item Type-$Q_1$ flipper: a flipper embedded in the first quadrant
  \item Type-$Q_2$ flipper: a flipper embedded in the second quadrant
  \item Type-$Q_4$ flipper: a flipper embedded in the fourth quadrant
  \item a variable box
  \item a clause box
\end{enumerate}
It is easy to check that each gadget belongs the exactly one
of these types. By construction, no two gadgets of the same
type may intersect. Take a variable box $X_i$ and a 
Type-$Q_2$ flipper $f_j$ joined to a variable box
$X_j$. Since $f_j$ is embedded strictly below the variable box
$X_j$ and strictly above the variable box $X_{j+1}$, $X_i$ and
$f_j$ cannot intersect, regardless of values for $i$ and $j$.

Now, take a clause box $C_i$ and a Type-$Q_4$ flipper $f_j$
joined to a clause box $C_j$. Observe that the grid 
line $x=(K+10a+2)n$ separates the embedding of flippers 
and clause boxes. Hence, $C_i$ and $f_j$ may not intersect,
regardless of values for $i$ and $j$.

For any other pair of types, they may not intersect since 
they belong to different quadrants. 
\qed
\end{proof}

\end{document}